\documentstyle[12pt]{article}
\begin{document}

\begin{center}

\vspace{5mm}

{\large \bf NOVEL FEATURES OF MULTIPLICITY DISTRIBUTIONS IN QCD AND EXPERIMENT}

\vspace{3mm}

{\large  I.M.\,DREMIN}

\vspace{2mm}

{\normalsize  Lebedev Physical Institute, Moscow 117924, Russia }

\vspace{3mm}

{\normalsize CONTENTS \\}

\end{center}

\noindent 1. Introduction \\
2. {\it Oscillations} of cumulants of multiplicity distributions in QCD \\
3. {\it Evolution} of distributions with decreasing phase space volume -- \\
intermittency and fractality \\
4. {\it Zeros} of truncated generating functions \\
5. Discussion and conclusions \\

\begin{abstract}
The solution of QCD equations for generating functions of multiplicity
distributions reveals new peculiar features of cumulant moments oscillating
as functions of their rank. This prediction is supported by experimental
data on $e^{+}e^{-}, hh, AA$ collisions. Evolution of the moments at smaller
phase space bins leads to intermittency and fractality. The experimentally
defined truncated generating functions possess zeros in the complex plane of
an auxiliary variable recalling Lee-Yang zeros in statistical mechanics.
\end{abstract}

\section{Introduction}

For a long time, the phenomenological approach dominated in description of
multiplicity distributions in multiparticle production. The very first
attempts to apply QCD formalism to the problem failed because in the
simplest double-logarithmic approximation it predicts an extremely wide shape
of the distribution that contradicts to
experimental data. Only recently it became possible to get exact solutions
of QCD equations which revealed much narrower shapes and such a novel feature
of cumulant moments as their oscillations at higher ranks. These moments are
extremely sensitive to the tiny details of the distribution. Surprisingly
enough, those QCD predictions for parton distributions have been supported
by experimental data for hadrons.

QCD is also successful in qualitative description of evolution of these
distributions with decreasing phase space bins which gives rise to notions
of intermittency and fractality. However, there are some new problems
with locations of the minimum of cumulants at small bins.

The experimentally defined truncated generating functions possess an intriguing
pattern of zeros in the complex plane of an auxiliary variable. It recalls
the pattern of Lee-Yang zeros of the grand canonical partition function in
the complex fugacity plane related to phase transition.

Before demonstrating all these peculiarities let us define the multiplicity
distribution
\begin{equation}
P_n = \sigma _n /\sum_{n=0}^{\infty }\sigma _n  ,     \label{1}
\end{equation}
where $\sigma _n$ is the cross section of $n$-particle production processes,
and the generating function
\begin{equation}
G(z) = \sum _{n=0}^{\infty }P_{n}(1+z)^n .     \label{2}
\end{equation}
The (normalized) factorial and cumulant moments of the $P_n$ distribution are
\begin{equation}
F_{q} =\frac {\sum_{n} P_{n} n(n-1)...(n-q+1)}{(\sum_{n} P_{n} n)^q} = \frac
{1}
{\langle n\rangle ^q} \frac {d^qG(z)}{dz^q}\vert _{z=0} ,  \label{3}
\end{equation}
\begin{equation}
K_{q} = \frac {1}{\langle n\rangle ^q}\frac {d^q \ln G(z)}{dz^q}\vert _{z=0},
\label{4}
\end{equation}
where $\langle n\rangle = \sum _{n} P_{n} n$ is the average multiplicity. They
describe full and genuine $q$-particle correlations, correspondingly. Let us
point out here that the moments are defined by the derivatives at the origin
and are very sensitive to any nearby singularity of the generating function.

In practice, one deals with distribution truncated due to finiteness of the
available phase space and the summation in all formulae above is cut off at
some finite value of $n=N_{max}$ which depends on the phase space region
chosen, and increases with its increase. It
is a polynomial of the power $N_{max}$ and has $N_{max}$ zeros
in the complex $z$-plane.

To shorten the presentation, I omit here all the details of calculations and
references to original papers. The reader can find them in my review paper in
Physics-Uspekhi {\bf 37} (1994) 715. Main qualitative results are described
and demonstrated in Figures in the subsequent three Sections. Their physics
implications are discussed in the last Section.

\section{Oscillations of cumulants of multiplicity distributions in QCD }

First, let us consider QCD without quarks, i.e. gluodynamics. The generating
function of the gluon multiplicity distribution in the full phase-space
volume satisfies the equation
\begin{equation}
\frac {\partial G(z,Y)}{\partial Y} = \int _{0}^{1}dxK(x)\gamma _{0}^{2}
[G(z,Y+\ln x)G(z,Y+\ln (1-x)) - G(z,Y)] .    \label{8}
\end{equation}
Here $Y=\ln (p\theta /Q_{0}), p$ is the initial momentum, $\theta $ is the
angular width of the gluon jet considered,  $p\theta \equiv Q$
where $Q$ is the jet virtuality, $Q_{0}=$const,
\begin{equation}
\gamma _{0}^{2} = \frac {6\alpha _{S}(Q)}{\pi } ,  \label{9}
\end{equation}
$\alpha _S$ is the running coupling constant, and the kernel of the equation is
\begin{equation}
K(x) = \frac {1}{x} - (1-x)[2-x(1-x)] .    \label{10}
\end{equation}
It is the non-linear integro-differential equation with shifted arguments in
the non-linear part which take into account the conservation laws, and with
the initial condition
\begin{equation}
G(z, Y=0) = 1+z ,    \label{11}
\end{equation}
and the normalization
\begin{equation}
G(z=0, Y) = 1 .     \label{12}
\end{equation}
The condition (\ref{12}) normalizes the total probability to 1, and the
condition (\ref{11}) declares that there is a single particle at the very
initial stage.

After Taylor series expansion at large enough $Y$ and differentiation in
eq. (\ref{8}), one gets the differential equation
\begin{equation}
(\ln G(Y))^{\prime \prime }= \gamma _{0}^{2}[G(Y)-1-2h_{1}G^{\prime }(Y)+
h_{2}G^{\prime \prime }(Y)] ,    \label{14}
\end{equation}
where $h_1 = 11/24; h_2 = (67-6\pi ^2)/36\approx 0.216$, and higher order terms
have been omitted.

Leaving two terms on the right-hand side, one gets the well-known equation
of the double-logarithmic approximation which takes into account the most
singular components. The next term, with $h_1$, corresponds to the modified
leading-logarithm approximation, and the term with $h_2$ deals with
next-to-leading corrections.

The straightforward solution of this equation looks very problematic.
However, it is very simple for the moments of the distribution because
$G(z)$ and $\ln G(z)$ are the generating functions of $F_q$ and $K_q$,
correspondingly, according to (\ref{3}), (\ref{4}). Using this fact, one gets
the solution which looks like
\begin{equation}
H_q = \frac {K_q}{F_q} = \frac {\gamma _{0}^{2}[1-2h_{1}q\gamma +h_{2}(q^2
\gamma ^{2} + q\gamma ^{\prime })]}{q^2 \gamma ^2 + q\gamma ^{\prime }},
\label{13}
\end{equation}
where the anomalous dimension $\gamma $ is related to
$\gamma _0$ by
\begin{equation}
\gamma = \gamma _0 - \frac {1}{2}h_{1}\gamma
_{0}^{2} + \frac {1}{8} (4h_2 - h_{1}^{2})\gamma _{0}^{3} + O(\gamma _{0}^{4})
.   \label{14}
\end{equation}

The formula (\ref{13}) shows how the ratio $H_q$ behaves in different
approximations. In double-log approximation when $h_1 = h_2 = 0$, it
monotonously decreases as $q^{-2}$ that corresponds to the negative binomial
law with its parameter $k=2$ i.e. to very wide distribution. In modified-log
approximation ($h_2 = 0$) it acquires a negative minimum at
\begin{equation}
q_{min}=\frac {1}{h_{1}\gamma _{0}}+\frac {1}{2}+O(\gamma _0) \approx 5
\label{q}
\end{equation}
  and approaches
asymptotically at large ranks $q$ the abscissa axis from below. In the next
approximation given by (\ref{13}) it preserves the minimum location
but approaches a
positive constant crossing the abscissa axis. In ever higher orders it
reveals the quasi-oscillatory behavior about this axis. This prediction of the
minimum at $q\approx 5$ and subsequent specific oscillations is the main
theoretical outcome.

It is interesting to note that the equation (\ref{8}) can be solved exactly
in the case of fixed coupling constant. All the above qualitative features
are noticeable here as well.

While the above results are valid for gluon distributions in gluon jets
(and pertain to QCD with quarks taken into account),
the similar qualitative features characterize the multiplicity distributions
of hadrons in high energy reactions initiated by various particles. As an
example, I show in Fig.1 the ratio $H_q$ as a function of $q$ in the
$e^{+}e^{-}$ data of DELPHI collaboration at 91 GeV, where the oscillations
and the location of minima are of a special interest.

\section{Evolution of distributions with decreasing phase-space volume --
intermittency and fractality}

The multiplicity distributions can be measured not only in the total phase
space (as has been discussed above for very large phase-space volumes) but
in any part of it. For the homogeneous distribution of particles within the
volume, the average multiplicity is proportional to the volume and decreases
for  small volumes but the fluctuations increase. The most interesting
problem here is the law governing the growth of fluctuations and its possible
departure from a purely statistical behavior related to the decrease of the
average multiplicity. Such a variation has to be connected with the dynamics of
the
interactions. In particular, it has been proposed to look for the power-law
behavior of the factorial moments for small rapidity intervals $\delta y$
\begin{equation}
F_q \propto (\delta y)^{-\phi (q)} \;\;\;\;(\phi (q)>0)\;\;\;\;
(\delta y \rightarrow 0), \label{15}
\end{equation}
inspired by the idea of intermittency in turbulence. In the case of statistical
fluctuations with purely Poissonian behavior, the intermittency indices $\phi
(q)$ are identically equal to zero.

Experimental data on various processes in a wide energy range support this idea
(e.g., see Fig.2), and QCD provides a good basis for its explanation as a
result
of parton showers. The generating function technique is not applicable here,
and
one should consider the Feynman graphs of evolution of a jet with its subjet
hitting the phase-space window under consideration (see the abovementioned
review).

At moderately small rapidity windows, one can get in the double-log
approximation the power-law behavior with
\begin{equation}
\phi (q) = D(q-1) - \frac {q^2 - 1}{q}\gamma _0 .   \label{16}
\end{equation}
The running property of QCD coupling constant is not important in that region.
This property becomes noticeable at ever smaller windows when (e.g., at $q$=2)
$\ln \delta y_0/\delta y >\alpha _{S}^{-1}$, and leads to smaller numerical
values of $\phi (q)$ compared to (\ref{16}). The general trends in this region
decline somewhat from the simple power law (\ref{15}) due to logarithmic
corrections. Qualitatively, these predictions correspond to experimental
findings at relatively small ranks $q$ where the steep increase
in the region of $\delta y>1$
on the log-log plot of the dependence (\ref{15}) is replaced by slower one at
smaller intervals $\delta y$ (see Fig.2). The transition point between
the two regimes depends on the rank in qualitative agreement with QCD
predictions also. Namely, the transition happens at smaller bins for higher
ranks. These findings can be interpreted as an indication on
fractal structure of particle distributions within the available phase space.
When interpreted in terms of fluctuations, they show that the fluctuations
become stronger in small phase-space regions in a definite power-like
manner and, surely, exceed trivial statistical fluctuations.

Let us turn now to the $q$-behavior of moments at small bins. The phenomenon of
the oscillations of cumulants discussed above reveals itself here as well if
one
goes beyond the double-log approximation of (\ref{16}). In terms of factorial
moments, it means the non-monotonous behavior of the intermittency indices as
functions of $q$. (Compare it to the steady increase with $q$ at $q>1$
given by (\ref{16}).) It gives rise to the negative values of $K_q$ and $H_q$.

The fate of the first minimum can be easily guessed from the formula (\ref{q}).
For large enough virtualities (i.e. small $\gamma _{0}$), the minimum location
moves to higher values of rank $q$ for jets
with larger virtuality $Q$ since the QCD coupling constant is running as
$\ln ^{-1}Q$. Therefore, the predicted shift of the minimum is
\begin{equation}
q_{min}\propto \ln ^{1/2}Q .
\end{equation}
It follows that $q_{min}$ moves to higher ranks at higher energies because
more massive jets become available. Another corollary is that it should
shift to smaller values of $q$ for smaller bins at fixed energy.

While former statement finds some support in experiment, the second one does
not look to be true as shown in Fig.3. On the contrary, the minimum appears
at higher ranks for smaller bins. There is no solution of this problem yet
but it should be ascribed to the higher-order effects.
Actually, one can guess that the higher order terms shown as $O(\gamma _{0})$
in (\ref{q}) become so important at small bins that they overpower the weak
$Q$-dependence of $\gamma _{0}^{-1}$ in the first term of (\ref{q}). It is
important to stress here that at large rapidity intervals the modified
leading-log term with $h_2$ does not influence the value of $q_{min}$, and
only increases the value of $H_q$ by $h_{2}\gamma _{0}^{2}$. Thus, the next-
to-leading corrections should be in charge of the additional shift of
$q_{min}$,
and, therefore, small bins help us look into higher orders of QCD.

\section{Zeros of truncated generating functions}

There is another fascinating feature of multiplicity distributions -- it
happens
that zeros of the truncated generating function form a spectacular pattern in
the complex plane of the variable $z$. Namely, they seem to lie close to a
single circle. At enlarged values of $N_{max}$ they move closer to the real
axis pinching it at some positive value of $z$. It is demonstrated in Fig.4
for UA5 data on $p\bar p$ interactions at 200 and 900 GeV for various
rapidity windows.

No QCD interpretation of the fact exists because it is hard to exploit the
finite cut-off in analytic calculations. The interest to it stems from the
analogy to the locations of zeros of the grand canonical partition function
as described
by Lee and Yang who related them to possible phase transitions in statistical
mechanics. In that case, $z$ variable plays the role of fugacity, and pinching
of the real axis implies existence of two phases in the system considered.

In particle physics, it shows up the location of the singularity of the
generating function i.e. the number of zeros of truncated generating functions
increases and they tend to move to the singularity point when $N_{max}
\rightarrow \infty $. Since it happens to lie close to the origin, it
drastically influences the behavior of moments (see (\ref{3}), (\ref{4})), and,
therefore, determines the shape of the distribution. The study of the
singularities is at the very early stage now, and one can only say that the
singularity is positioned closer to the origin in nucleus-nucleus collisions
and it is farthest in $e^{+}e^{-}$ that appeals to our intuitive guess.

\section{Discussion and conclusions}

Let us discuss the implications of the above findings. The QCD prediction
of quasioscillating behavior of cumulant moments of the multiplicity
distributions reveals the tiny features which were overlooked by simpleminded
fits of the negative binomial distribution. The truncated negative binomial
distribution has the oscillating cumulants as well but the oscillations should
die out asymptotically while they persist in QCD.

It demonstrates that the QCD distribution belongs to the class
of non-infinitely-divisible ones and shows that the Poissonian
cluster models are ruled out by QCD since they lead to positive
values of cumulants and can not reproduce their oscillations at
asymptotically high energies.

Moreover, this prediction is closely related to the existence
of a new expansion parameter in describing multiparticle production
in QCD. In terms of moments, this parameter is equal to the product of the
moment's rank to the anomalous dimension of QCD $\gamma q$. Let us
recall that the similar parameter appears when calculating the Feynman tree
graphs and looks like a product of the number of final particles to QCD
coupling constant $n\alpha _S$. Both parameters become large if large
number of particles is involved.

The invalidity of simple perturbative approach in multiple production processes
would ask for more convenient basis (compared to particle number
representation)
to be used. As one of possible examples of effectiveness of such an approach,
we would mention lasers where coherent state basis is more suitable. That is
why studies of coherent and squeezed states as well as statistical approach
and search for collective effects, in general, are welcome. The recent
findings of zeros of the truncated generating functions point out in that
direction, and provoke speculations to their possible relation to the problem
of phase transition in hadronic matter.

The success of QCD in predicting the qualitative features of moments of
multiplicity distributions both for large phase-space regions (oscillation
of cumulants) and for small bins (intermittency indices, fractality) seems
even more surprising if one recognizes that they are derived for parton
distributions while hadrons are observed in experiment. Nevertheless, these
qualitative features reveal themselves in hadron distributions as well.
Why such a local parton-hadron duality persists even at higher moments is
an open question. Another problem appears in such a tiny property as
the evolution of the minimum location at smaller bins. Its solution can show
a way to take into account higher order effects of QCD more properly.

To conclude, we would like to stress that recent theoretical studies of QCD
multiplicity distributions predict rather exotic features of the moments of the
distributions which find some support in experiment and, at the same time,
provoke new problems to be solved.

\vspace{1mm}

{\large Acknowledgements}

\vspace{1mm}

I am grateful to Takeshi Kodama for inviting me to participate in this very
fruitful workshop and for financial support.

This work is supported by the Russian Fund for Fundamental Research (grant
93-02-3815), by Soros Foundation (grant M5V300) and by INTAS-930-0079 grant.\\

\begin{center}

FIGURE CAPTIONS

\end{center}

\vspace{2mm}

Fig.1 The ratio of cumulant to factorial moments $H_q$ as the function of the
rank $q$. The DELPHI-data on $e^{+}e^{-}$ at 91 GeV in the total phase space
are shown by dots. The dashed line shows the fit by the negative binomial
distribution with the parameters given by DELPHI. The solid line is drawn just
to guide the eye.

\vspace{1mm}

Fig.2 Factorial moments of order $q$=2,$\ldots $,4 for the all-charged,
positives-only and negatives-only samples
of NA22-data on $\pi ^{+}p$ interactions at 250 GeV/c (lab. system).

\vspace{1mm}

Fig.3 Behavior of the ratio of cumulant to factorial moments $H_q$ as a
function
of the rank $q$ for some single hemisphere multiplicity distributions from
the DELPHI collaboration: (a) full hemisphere; (b) rapidity window
$\vert y\vert \leq 1.5$; (c) rapidity window $\vert y\vert \leq 1.0$.
The minimum shifts to higher $q$ for smaller bins.

\vspace{1mm}

Fig.4 The locations of zeros of the truncated generating function for UA5-data
on $p\bar p$ interactions at 200 and 900 GeV (c.m.s.) in various rapidity
windows. The upper halfplane of $z=x+iy$ is shown only because of the
up-down symmetry. (For further experimental information see the talk of
G. Gianini at Multiparticle Dynamics-95).

\end{document}